\documentclass{article}
\usepackage{amssymb}
\usepackage{amsfonts}
\usepackage{amsmath}

\newtheorem{theorem}{Theorem}
\newtheorem{acknowledgement}[theorem]{Acknowledgement}

\input{tcilatex}

\begin{document}

\title{Melnikov method approach to control of homoclinic/heteroclinic chaos
by weak harmonic excitations}
\author{\textsc{Ricardo Chacon} \\
%EndAName
\textit{Departamento de Electr\'{o}nica e Ingenier\'{\i}a Electromec\'{a}%
nica,}\\
\textit{\ Escuela de Ingenier\'{\i}as Industriales, Universidad de
Extremadura, }\\
\textit{Apartado Postal 382, E-06071 Badajoz, Spain}}
\maketitle

\begin{abstract}
A review on the application of Melnikov%
%TCIMACRO{\U{b4}}%
%BeginExpansion
\'{}%
%EndExpansion
s method to control homoclinic and heteroclinic chaos in low-dimensional,
non-autonomous and dissipative, oscillator systems by weak harmonic
excitations is presented, including diverse applications such as chaotic
escape from a potential well, chaotic solitons in Frenkel-Kontorova chains,
and chaotic charged particles in the field of an electrostatic wave packet.

\end{abstract}

\section{Introduction}

During the past 15 years or so, diverse techniques of non-feedback chaos
control have been proposed (Chen \& Dong 1998) that may be roughly
classified into three types: (i) The parametric excitation of an
experimentally adjustable parameter; (ii) Entrainment to the target
dynamics; and (iii) The application of a coordinate-independent (or dipole)
external periodic excitation. It is shown below that techniques (i) and
(iii) may be unified in a general setting for the class of dissipative
systems considered in this present review. There exists numerical,
experimental, and theoretical evidence that the period of the most effective
chaos-controlling excitations usually is a rational fraction of a certain
period associated with the uncontrolled system, although the effectiveness
of incommensurate excitations has also been demonstrated in some particular
cases (Chac\'{o}n \& Mart\'{\i}nez 2002). Indeed, resonances between the
chaos-controlling excitation and (i) a periodic chaos-inducing excitation,
(ii) an unstable periodic orbit embedded in the chaotic attractor, (iii) a
natural period in a period doubling route to chaos, or (iv) a period
associated with some peak in the power spectrum, have been considered in
diverse successful chaos-controlling excitations. This is not really
surprising since these types of resonances are closely related to each
other. For instance, when a damped, harmonically forced oscillator exhibits
a steady chaotic state, the power spectrum corresponding to a given state
variable typically presents its main peaks at frequencies which are rational
fractions of the chaos-inducing frequency for certain ranges of the
chaos-inducing amplitude. The extensive literature concerning experimental,
theoretical, and numerical studies of non-feedback methods is frankly
unapproachable because of its volume in a review of the present type.
Therefore, only pioneering key work (from the author%
%TCIMACRO{\U{b4}}%
%BeginExpansion
\'{}%
%EndExpansion
s viewpoint) is mentioned in the following. The effectiveness of periodic
parametric excitations in suppressing chaos was shown by Alekseev \&
Loskutov (1987). H\"{u}bler \& L\"{u}scher (1989) discussed how a nonlinear
oscillator can be driven toward a given target dynamics by means of resonant
excitations. Braiman \& Goldhirsch (1991) provided numerical evidence to
show the possibility of inhibiting chaos by an additional periodic
coordinate-independent excitation. Salerno (1991) showed, by the analysis of
a phase-locked map, the possibility of suppressing chaos in long
biharmonically driven Josephson junctions. Chac\'{o}n \& D\'{\i}az Bejarano
(1993) discussed a new way to reduce or suppress steady chaotic states, by
only altering the geometrical shape of weak periodic perturbations. Kivshar 
\textit{et al}. (1994) \ showed analytically and numerically that the
suppression of chaos may be effectively achieved by applying a
high-frequency parametric force to a chaotic dynamical system. Experimental
control of chaos by weak periodic excitations has been demonstrated in many
diverse systems, including magnetoelastic systems (Ditto \textit{et al.}
1990), ferromagnetic systems (Azevedo \& Rezende 1991), electronic systems
(Hunt 1991), laser systems (Roy \textit{et al.} 1992; Meucci \textit{et al}.
1994; Chizhevsky \& Corbal\'{a}n 1996; Uchida \textit{et al. }1998),
chemical reactions (Petrov \textit{et al.} 1993; Alonso \textit{et al.}
2003), neurological systems (Schiff \textit{et al.} 1994), and plasma
systems (Ding \textit{et al.} 1994).

This paper summarizes some main results concerning the application of
Melnikov%
%TCIMACRO{\U{b4}}%
%BeginExpansion
\'{}%
%EndExpansion
s method (Melnikov 1963; Arnold 1964; Guckenheimer \& Holmes 1983; Wiggings
1990) to the problem of control of chaos in low-dimensional, non-autonomous
and dissipative, oscillator systems by small-amplitude harmonic
perturbations. Specifically, the class of systems considered is described by
the differential equation%
\begin{equation}
\overset{..}{x}+\frac{dU(x)}{dx}=-d(x,\overset{.}{x})+p_{c}(x,\overset{.}{x}%
)F_{c}(t)+p_{s}\left( x,\overset{.}{x}\right) F_{s}(t),  \tag{1.1}
\end{equation}%
where $U(x)$ is a nonlinear potential, $-d(x,\overset{.}{x})$ is a generic
dissipative force which may include constant forces and time-delay terms, $%
p_{c}(x,\overset{.}{x})F_{c}(t)$ is a chaos-inducing excitation, and $%
p_{s}\left( x,\overset{.}{x}\right) F_{s}(t)$ is an as yet undetermined
suitable chaos-controlling excitation, with $F_{c}(t)$, $F_{s}(t)$ being
harmonic functions of initial phases $0,\Theta $, and frequencies $\omega
,\Omega $, respectively. It is worth mentioning that Melnikov%
%TCIMACRO{\U{b4}}%
%BeginExpansion
\'{}%
%EndExpansion
s method imposes on (1.1) some additional limitations: the excitation,
time-delay, and dissipation terms are weak perturbations of the underlying
conservative system $\overset{..}{x}+dU(x)/dx=0$ which has a separatrix. The
original work of Melnikov (1963) was generalized by Arnold (1964) to a
particular instance of a time-periodic Hamiltonian perturbation of a
two-degree-of-freedom integrable Hamiltonian system. Fifteen years later,
Holmes (1979) was the first to apply Melnikov%
%TCIMACRO{\U{b4}}%
%BeginExpansion
\'{}%
%EndExpansion
s method (to a damped forced two-well Duffing oscillator) in the west. From
then on the method began to be popular. Chow \textit{et al.} (1980)
rediscovered Melnikov%
%TCIMACRO{\U{b4}}%
%BeginExpansion
\'{}%
%EndExpansion
s results using alternative methods and emphasized that homoclinic and
subharmonic bifurcations are closely related. Through the 1980s a great
variety of extensions and generalizations of Melnikov%
%TCIMACRO{\U{b4}}%
%BeginExpansion
\'{}%
%EndExpansion
s approach were developed (Greenspan 1981; Holmes \& Marsden 1982; Lerman \&
Umanski 1984; Greundler 1985; Salam 1987; Schecter 1987; Wiggings 1987). The
interested reader is referred to the books by Lichtenberg \& Lieberman
(1983), Guckenheimer \& Holmes (1983), Wiggings (1988), and Arrowsmith \&
Place (1990) for more details and references. The work of Cai \textit{et al.}
(2002) provides the simplest extension of Melnikov%
%TCIMACRO{\U{b4}}%
%BeginExpansion
\'{}%
%EndExpansion
s method to include perturbational time-delay terms.

The application of Melnikov%
%TCIMACRO{\U{b4}}%
%BeginExpansion
\'{}%
%EndExpansion
s method to controlling chaos in low-dimensional systems by weak periodic
perturbations began in about 1990. Indeed, Lima \& Pettini (1990) provided a
heuristic argument to extend the idea that parametric perturbations can
modify the stability of hyperbolic or elliptic fixed points, in the phase
space of linear systems, to the case of nonlinear systems, and hence that
parametric perturbations could also provide a means to reduce or suppress
chaos in nonlinear systems. They used for the first time the Melnikov method
to analytically demonstrate this conjecture in the case of a damped driven
two-well Duffing oscillator subjected to a chaos-suppressing parametric
excitation. However, their insufficient analysis of the corresponding
Melnikov function led them into gross errors in their final results and
conclusions. Specifically, they failed both theoretically to demonstrate the
sensitivity of the suppression scenario to the initial phase of the
chaos-suppressing excitation and to find it numerically. They also failed
theoretically to predict the suppression of chaos in the case of subharmonic
resonances (between the chaos-inducing and chaos-suppressing excitations)
higher than the main one. Although a part of their erroneous analysis of the
Melnikov function originated from a mistake in its calculation (Cuadros \&
Chac\'{o}n 1993; Lima \& Pettini 1993), its main weakness was in not
providing a correct necessary and \textit{sufficient} condition for the
Melnikov function to always have the same sign (i.e., for the frustration of
homoclinic bifurcations). For the two-well Duffing oscillator that they
considered, such a correct necessary and sufficient condition was first
deduced for the general case of subharmonic resonances by Chac\'{o}n (1995%
\textit{a}), where the extremely important role of the initial phase (of the
chaos-suppressing excitation) on the suppression scenario was demonstrated
theoretically. Cicogna \& Fronzoni (1990) studied the suppression of chaos
in the Josephson-junction model%
\begin{equation}
\overset{..}{\phi }+\left[ 1+\xi \cos \left( \Omega t+\theta \right) \right]
\sin \phi =-\delta \overset{.}{\phi }+\gamma \cos \left( \omega t\right) , 
\tag{1.2}
\end{equation}%
where the parametric excitation $\xi \cos \left( \Omega t+\theta \right)
\sin \phi $ is the chaos-suppressing excitation, for the single case of the
main resonance $\Omega =\omega $ by using Melnikov%
%TCIMACRO{\U{b4}}%
%BeginExpansion
\'{}%
%EndExpansion
s method. Their insufficient analysis of the Melnikov function (in
particular, that of the role played by the initial phase $\theta $) led them
also into gross mistakes in their final conclusions. On the contrary, it was
demonstrated by Chac\'{o}n (1995\textit{b}) that the effect of the above
parametric excitation in (1.2) for the general case of subharmonic
resonances ($\Omega =p\omega $, $p$ an integer) is to suppress the chaotic
behavior when a \textit{suitable} initial phase is used and only for \textit{%
certain} ranges of its amplitude. It was also shown (Chac\'{o}n 1995\textit{b%
}) for the first time that such suitable initial phases are compatible with
the surviving natural symmetry under the parametric excitation. It was also
conjectured (Chac\'{o}n 1998) that such maximum survival of the symmetries
of solutions from a broad and relevant class of systems, subjected both to
primary chaos-inducing and chaos-suppressing excitations, corresponds to the
optimal choice of the suppressory parameters; specifically, to particular
values of the initial phase differences between the two types of excitations
for which the amplitude range of the suppressory excitation is maximum.
Rajasekar (1993) applied Melnikov%
%TCIMACRO{\U{b4}}%
%BeginExpansion
\'{}%
%EndExpansion
s method to study the suppression of chaos in the Duffing-van der Pol
oscillator%
\begin{equation}
\overset{..}{x}-\alpha ^{2}x+\beta x^{3}=-p\left( 1-x^{2}\right) \overset{.}{%
x}+f\cos \left( \omega t\right) +\eta \cos \left( \Omega t+\Omega \phi
\right) ,  \tag{1.3}
\end{equation}%
where the additional forcing $\eta \cos \left( \Omega t+\Omega \phi \right) $
is the chaos-suppressing excitation, for the single case of the main
resonance $\Omega =\omega $. He pointed out the relevant role of the initial
phase (of the chaos-suppressing excitation) on the suppression scenario for
the first time, and he also deduced the analytical expression of the
boundaries of the regions in the $\eta -\phi $ phase plane where homoclinic
chaos is inhibited. A generalization of Rajasekar%
%TCIMACRO{\U{b4}}%
%BeginExpansion
\'{}%
%EndExpansion
s approach concerning the relative effectiveness of any two weak excitations
in suppressing homoclinic/heteroclinic chaos is discussed in the work of Chac%
\'{o}n (2002). There, general analytical expressions are derived from the
analysis of generic Melnikov functions providing the boundaries of the
regions as well as the enclosed area in the amplitude-initial phase plane of
the chaos-suppressing excitation where homoclinic/heteroclinic chaos is
inhibited. Also, a criterion based on the aforementioned area was deduced
and shown to be useful \ in choosing the most suitable of the possible
chaos-suppressing excitations. Cicogna \& Fronzoni (1993) analyzed the
Melnikov function associated with the family of systems 
\begin{equation}
\overset{..}{x}=f(x)-\delta \overset{.}{x}+\gamma \cos \left( \omega
t\right) +\varepsilon g\left( x\right) \cos \left( \Omega t+\theta \right) ,
\tag{1.4}
\end{equation}%
where $\varepsilon g\left( x\right) \cos \left( \Omega t+\theta \right) $ is
the chaos-suppressing excitation, for the single case of the main resonance $%
\Omega =\omega $. They deduced both the suitable suppressory values of the
initial phase $\theta $ and the associated chaotic threshold function $%
\left( \gamma /\delta \right) _{th}$ when the chaos-suppressing excitation
acts on the system. General results (Chac\'{o}n 1999) concerning suppression
of homoclinic/heteroclinic chaos were derived on the basis of Melnikov%
%TCIMACRO{\U{b4}}%
%BeginExpansion
\'{}%
%EndExpansion
s for the family (1.1) for the general case of subharmonic resonance ($%
\Omega =p\omega $, $p$ an integer). There, a generic analytical expression
was deduced for the maximum width of the intervals of the initial phase $%
\Theta $ for which homoclinic/heteroclinic bifurcations can be frustrated.
It was also demonstrated that $\left\{ 0,\pi /2,\pi ,3\pi /2\right\} $ are,
in general, the only optimal values of such initial phase, in the sense that
they allow the widest amplitude ranges for the chaos-suppressing excitation.
The work of Chac\'{o}n (2001\textit{a}) presents general results concerning 
\textit{enhancement }or maintenance of chaos for the family (1.1), where the
connection with the results on chaos suppression was discussed in a general
setting. It was also demonstrated that, in general, a second harmonic
excitation can reliably play an enhancer or inhibitor role by solely
adjusting its initial phase. The work of Chac\'{o}n (2001\textit{b})
provides a preliminary Melnikov-method-based approach concerning suppression
of chaos by a chaos-suppressing excitation which satisfies an \textit{%
ultrasubharmonic} resonance condition with the chaos-inducing excitation.
This approach was further applied to the problem of the inhibition of
chaotic escape from a potential well by \textit{incommensurate}
escape-suppressing excitations (Chac\'{o}n \& Mart\'{\i}nez 2002).

\section{Basic theoretical approach}

To illustrate the theoretical approach with a paradigmatic example, consider
a single Josephson junction subjected to a nonlinear dissipative term and
driven by two harmonic excitations (Chac\'{o}n \textit{et al.} 2001)%
\begin{equation}
\overset{..}{x}+\sin x=-\alpha (1+\gamma \cos x)\overset{.}{x}+F\sin \left(
\omega t\right) +\beta F\sin \left( \Omega t+\Theta \right) ,  \tag{2.1}
\end{equation}%
where $x$ and time are dimensionless variables, and $\overset{.}{x}$ is
proportional to the difference of potential between the two superconductors.
It is also assumed that the terms of dissipation and excitation are regarded
as weak perturbations and $\beta F\sin \left( \Omega t+\Psi \right) $ is the
chaos-suppressing excitation. The nonlinear dissipative term appears in the
study of a single Josephson junction when the conditions are such that the
interference effects between the pair and quasiparticle currents should be
taken into account (Barone \& Paterno 1982). The application of the Melnikov
method to (2.1) yields the Melnikov function%
\begin{equation}
M^{\pm }\left( t_{0}\right) =-C\pm A\sin \left( \omega t_{0}\right) \pm
B\sin \left( \Omega t_{0}+\Theta \right) ,  \tag{2.2}
\end{equation}%
with%
\begin{eqnarray}
C &\equiv &8\alpha \left( 1+\gamma /3\right) ,  \notag \\
A &\equiv &2\pi F\func{sech}\left( \frac{\pi \omega }{2}\right) ,  \notag \\
B &\equiv &2\pi \beta \func{sech}\left( \frac{\pi \Omega }{2}\right) . 
\TCItag{2.3}
\end{eqnarray}%
Turning to the general case (1.1), let us assume that such a family of
systems satisfies the requirements of the Melnikov method. Then, the
application of the method to (1.1) provides the generic Melnikov function%
\begin{equation}
M_{h,h^{\prime }}^{\pm }\left( \tau _{0}\right) =D\pm Ahar\left( \omega \tau
_{0}\right) +Bhar^{\prime }\left( \Omega \tau _{0}+\Psi _{h,h^{\prime
}}^{\pm }\right) ,  \tag{2.4}
\end{equation}%
where $\limfunc{har}\left( \tau \right) $ means indistinctly $\sin \left(
\tau \right) $ or $\cos \left( \tau \right) $, and $A$ is a non-negative
function, while $D,B$ can be non-negative or negative functions, depending
upon the respective parameters for each specific system. In particular, $D$
contains the effect of the damping, time-delay terms, and constant forces.
In the absence of time-delay terms and constant drivings, $D<0$, while one
has the three cases $D\gtrless 0$ when a constant driving and a time-delay
term act on the system. Also, $A$ and $B$ contain the effect of the
chaos-inducing and chaos-controlling excitations, respectively. Note that
changing the sign of $B$ is equivalent to having a fixed shift of the
initial phase: $B\rightarrow -B\Longleftrightarrow \Psi _{h,h^{\prime
}}^{\pm }\rightarrow $ $\Psi _{h,h^{\prime }}^{\pm }\pm \pi $, where the two
signs before $\pi $ apply to each of the sign superscripts of $\Psi $.
Therefore, $B$ will be considered (for example) as a positive function in
the following. As phase and initial time $\tau _{0}$ are not fixed, one may
study the simple zeros of $M_{h,h^{\prime }}^{\pm }\left( \tau _{0}\right) $
by choosing quite freely the trigonometric functions in (2.4). Therefore,
consider, for instant, the Josephson junction given by (2.1). It is worth
noting that the Melnikov functions $M_{h,h^{\prime }}^{\pm }\left( \tau
_{0}\right) $, $M^{\pm }\left( t_{0}\right) $ (cf. (2.4) and (2.2),
respectively) are connected by \textit{linear} relationships which are known
for each specific system (1.1):%
\begin{eqnarray}
\tau _{0} &=&\tau _{0}\left( t_{0},\omega \right) ,  \TCItag{2.5} \\
\Psi _{h,h^{\prime }}^{\pm } &=&\Psi _{h,h^{\prime }}^{\pm }\left( \Theta ,%
\frac{\Omega }{\omega }\right) .  \notag
\end{eqnarray}%
Therefore, the control theorems associated with any Melnikov function $%
M_{h,h^{\prime }}^{\pm }\left( \tau _{0}\right) $ can be straightforwardly
obtained from those associated with $M^{\pm }\left( t_{0}\right) $ (Chac\'{o}%
n 1999).

\subsection{Suppression of chaos}

As is well-known, the Melnikov method provides estimates in parameter space
for the appearance of homoclinic (and heteroclinic) bifurcations, and hence
for transient chaos. This means that in most of cases only \textit{necessary 
}conditions for steady chaos (strange chaotic attractor) are obtained from
the method. Therefore, one may always get sufficient conditions for the
inhibition of even transient chaos (frustration of homoclinic/heteroclinic
bifurcation) and, \textit{a fortiori}, for the inhibition of the steady
chaos that ultimately arises from such a homoclinic/heteroclinic
bifurcation. This is the principal foundation of the utility of Melnikov
method in predicting the suppression of (steady) chaos when a
homoclinic/heteroclinic bifurcation occurs prior to its emergence. For the
Josephson junction (2.1) one has the following theorem (Chac\'{o}n \textit{%
et al.} 2001):

\begin{theorem}
Let $\Omega =p\omega $, $p$ an integer, such that, for $M^{+}\left(
t_{0}\right) $ $\left( M^{-}\left( t_{0}\right) \right) $, $p=\frac{%
4m-1-2\Theta /\pi }{4n+1}$ $\left( p=\frac{4m+1-2\Theta /\pi }{4n-1}\right) $
is satisfied for some integers $m$ and $n$. Then $M^{\pm }\left(
t_{0}\right) $ always has the same sign, specifically $M^{\pm }\left(
t_{0}\right) <0$, if and only if the following condition is satisfied:%
\begin{eqnarray}
\beta _{\min } &<&\beta \leqslant \beta _{\max },  \notag \\
\beta _{\min } &\equiv &\left( 1-\frac{C}{A}\right) R,  \notag \\
\beta _{\max } &\equiv &\frac{R}{p^{2}},  \notag \\
R &\equiv &\frac{\cosh \left( \pi \Omega /2\right) }{\cosh \left( \pi \omega
/2\right) }.  \TCItag{2.6}
\end{eqnarray}
\end{theorem}

Now, the following remarks are in order.

First, one can test the suppression theorem theoretically by considering the
limiting Hamiltonian case $\left( \alpha =0\right) $. Notice that, in the
absence of dissipation, (2.6) must be rewritten as $\beta _{\min }\leqslant
\beta \leqslant \beta _{\max }$, $\beta _{\min }\equiv R$, $\beta _{\max
}\equiv R/p^{2}$, since $\beta _{\min }$ cannot now be zero. Thus, one
obtains (Chac\'{o}n \textit{et al.} 2001) $\Omega =\omega $, $\Theta =\pi $,
and $\beta =1$ as a necessary and sufficient condition for suppressing
stochasticity. (This result can be trivially obtained, to \textit{first
perturbative order}, from (2.2), (2.3) with $\alpha =0$, i.e., having $%
M^{\pm }\left( t_{0}\right) =0$ for all $t_{0}$.

Second, the lower threshold for the chaos-suppressing amplitude, $\beta
_{\min }$, takes into account the \textit{strength} of the initial chaotic
state through the factor $\left( 1-C/A\right) $, since one usually finds
that the corresponding maximal Lyapunov exponent $\lambda ^{+}$ increases as
the ratio $C/A$ decreases over a certain range of parameters. Therefore, for
fixed-chaos inducing and chaos-suppressing frequencies (and hence $R$
fixed), one would expect that $\beta _{\min }$ will increase as $\lambda
^{+} $ is increased. Note that the corresponding upper threshold, $\beta
_{\max }$, does not verify this important property, which is because $\beta
_{\max }$ arises from a \textit{necessary} condition for the necessary
condition yielding $\beta _{\min }$ to be also a sufficient condition. This
means that $\beta _{\max }$ slightly underestimates the upper threshold for
the chaos-suppressing amplitude, as is numerically and experimentally
observed in different instances. It is worth noting that this remark holds
for any Melnikov function (2.4).

Third, the asymmetry between the upper and lower homoclinic orbits (cf.
(2.2), (2.3)) gives rise to two distinct sets of \textit{optimal} initial
phases that are suitable for suppressing chaos. The optimal suppressory
values of $\Theta $ (hereafter denoted as $\Theta _{opt}$) are those values
allowing the widest amplitude ranges for the chaos-suppressing excitation
(the use of the adjective is justified below in the discussion of the
suitable intervals of initial phase difference for taming chaos). Indeed,
Theorem 1 requires having $\Theta =\Theta _{opt}\equiv \pi ,\pi /2,0,3\pi /2$
$\left( \pi ,3\pi /2,0,\pi /2\right) $ for $p=4n-3,4n-2,4n-1,4n$ $\left(
n=1,2,...\right) $, respectively, in order to inhibit chaos when one
considers orbits initiated near the upper (lower) homoclinic orbit. These
distinct values are those compatible with the \textit{surviving natural
symmetry} under the additional forcing. Indeed, the dissipative,
harmonically driven Josephson junction ($\beta =0$) is invariant under the
transformation 
\begin{eqnarray}
x &\rightarrow &-x,  \notag \\
t &\rightarrow &t+\frac{\left( 2n+1\right) \pi }{\omega },  \TCItag{2.7}
\end{eqnarray}%
where $n$ is an integer, i.e., if $\left[ x(t),\overset{.}{x}(t)\right] $ is
a solution of (2.1) with $\beta =0$, then so is $\left[ -x(t+\left(
2n+1\right) \pi /\omega ),-\overset{.}{x}(t+\left( 2n+1\right) \pi /\omega )%
\right] $. This pair of solutions may be essentially the same in the sense
that they may differ by an integer number of complete cycles, i.e., 
\begin{equation}
x(t)=-x\left[ t+\left( 2n+1\right) \pi /\omega \right] +2\pi l,  \tag{2.8}
\end{equation}%
with $l$ an integer, and they are termed symmetric. Otherwise, the
time-shifting and sign reversal procedure yields a different solution, and
the two solutions are termed broken-symmetric. When $\beta >0$ and $\Theta $
is arbitrary the aforementioned natural symmetry is generally broken. The
reason for that breaking is 
\begin{equation}
\sin \left( \Omega t+\Theta \right) \neq \sin \left[ \Omega t+\Theta +\left(
2n+1\right) \pi \Omega /\omega \right] ,  \tag{2.9}
\end{equation}%
for arbitrary $\omega ,\Omega ,$ and $\Theta $. Assuming a resonance
condition $\Omega =p\omega $, the survival of the above symmetry implies%
\begin{equation}
\sin \left( p\omega t+\Theta \right) =\left( -1\right) ^{p+1}\sin \left(
p\omega t+\Theta \right) .  \tag{2.10}
\end{equation}%
Obviously, this is only the case for $p$ an \textit{odd} integer. For $p$ an
even integer, one has the new transformation [$x\rightarrow -x$, $%
t\rightarrow t+\left( 2n+1\right) \pi /\omega $, $\Theta \rightarrow \Theta
\pm \pi $]. In other words, if $\left[ x(t),\overset{.}{x}(t)\right] $ is a
solution for a value $\Theta $, then so is $\left[ -x(t+\left( 2n+1\right)
\pi /\omega ),-\overset{.}{x}(t+\left( 2n+1\right) \pi /\omega )\right] $
for $\Theta \pm \pi $. Thus, this explains the origin of the differences
between the corresponding (at the same resonance order) allowed $\Theta
_{opt}$ values for the two homoclinic orbits. Similar results have been
found for the damped, driven pendulum mounted on a vertically oscillating
point of suspension (Chac\'{o}n 1995\textit{b}). Therefore, this maximum
symmetry principle appears to be the common background in the mechanism of
regularization by the application of resonant excitations.

Fourth, the width of the allowed interval $\left] \beta _{\min },\beta
_{\max }\right] $ for regularization is 
\begin{equation}
\Delta \beta \equiv \beta _{\max }-\beta _{\min }=\left[ \frac{C}{A}-\frac{%
p^{2}-1}{p^{2}}\right] R,  \tag{2.11}
\end{equation}%
with $R$ given by (2.6). Since $R$ is a positive function, there always
exists a \textit{maximum resonance order} $p_{\max }$ for suppression of
homoclinic (and heteroclinic) chaos, for each fixed initial chaotic state
(i.e., $C/A$ fixed), which is 
\begin{equation}
p_{\max }\equiv \left[ \left( 1-\frac{C}{A}\right) ^{-1/2}\right] , 
\tag{2.12}
\end{equation}%
where the brackets indicate integer part. From (2.12), one sees that $%
p_{\max }$ increases as the ratio $C/A$ is increased, which would associated
with the decrease of the corresponding maximal Lyapunov exponent over a
certain range of parameters. For a given set of parameters satisfying the
above theorem%
%TCIMACRO{\U{b4}}%
%BeginExpansion
\'{}%
%EndExpansion
s hypothesis, as the resonance order $p$ is increased the allowed interval $%
\left] \beta _{\min },\beta _{\max }\right] $ shrinks quickly for \textit{low%
} frequencies. This means that initial chaotic states cannot necessarily be
regularized to periodic attractors of arbitrary \textit{long} period, since
numerical experiments indicate that the regularized responses are typically
a period-1 attractor for $p=1$ and a period-2 attractor for $p=2$. On the
other hand, the asymptotic behavior $\Delta \beta \left( \omega \rightarrow
\infty \right) =\infty $ (the remaining parameters being held constant)
means that chaotic motion is not possible in this limit, as expected.

Fifth, to establish the suppression theorem corresponding to any Melnikov
function (2.4), it is enough to transform $M_{h,h^{\prime }}^{\pm }\left(
\tau _{0}\right) $ into the form given by (2.2). Therefore, taking into
account (2.5) and the aforementioned $\Theta _{opt}$ values, one finds that
in general there exist at most four suitable optimal values for the
suppressory initial phase difference between the two (commensurate: $\Omega
=p\omega $) excitations: $0,\pi /2,\pi ,3\pi /2$.

Sixth, It has been stated above that the suitable values of the initial
phase difference (between the two excitations involved) given by Theorem 1
are optimal, in the sense that they allow the widest amplitude ranges for
the chaos-suppressing excitation. One therefore could expect reliable
control of the dynamics over certain suitable phase difference intervals,
which would be centered on such optimal values, although this would imply a
reduction of the respective amplitude ranges. It has been deduced (Chac\'{o}%
n 1999; Chac\'{o}n \textit{et al.} 2001) that there always exists a \textit{%
maximum-range} interval 
\begin{equation}
\left[ \Theta _{opt}-\Delta \Theta _{\max },\Theta _{opt}+\Delta \Theta
_{\max }\right] ,  \tag{2.13}
\end{equation}%
of permitted initial phase differences for homoclinic/heteroclinic chaos
inhibition, where%
\begin{equation}
\Delta \Theta _{\max }\equiv \arcsin \left( \frac{C}{A}\right) .  \tag{2.14}
\end{equation}%
For each value of $\Theta $ belonging to this interval there exists a
reduced interval (with regard to the limiting case where the only suitable
values of $\Theta $ are $\Theta _{opt}$) of amplitudes of the
chaos-suppressing excitation which is%
\begin{eqnarray}
\beta _{\min }\left( \Theta =\Theta _{opt}\pm \Delta \Theta \right) &<&\beta
\leqslant \beta _{\max }\left( \Theta =\Theta _{opt}\pm \Delta \Theta
\right) ,  \notag \\
\beta _{\min }\left( \Theta =\Theta _{opt}\pm \Delta \Theta \right) &\equiv
&\left( 1-\frac{C}{A}\right) R\sec \left( \Delta \Theta \right) ,  \notag \\
\beta _{\max }\left( \Theta =\Theta _{opt}\pm \Delta \Theta \right) &\equiv &%
\frac{R\cos \left( \Delta \Theta \right) }{p^{2}},  \TCItag{2.15}
\end{eqnarray}%
where $R$ is given by (2.6) and $0\leqslant \Delta \Theta \leqslant \Delta
\Theta _{\max }$. Thus, the width of the range for the chaos-suppressing
amplitude is 
\begin{equation}
\Delta \beta \left( \Theta =\Theta _{opt}\pm \Delta \Theta \right) =\left\{ 
\frac{\cos \left( \Delta \Theta \right) }{p^{2}}-\frac{1-C/A}{\cos \left(
\Delta \Theta \right) }\right\} R,  \tag{2.16}
\end{equation}%
i.e., for fixed $C,A$ and $\Delta \Theta $ there always exists a \textit{%
maximum resonance order} $p_{\max }$ for homoclinic chaos suppression which
is%
\begin{equation}
p_{\max }\left( \Theta =\Theta _{opt}\pm \Delta \Theta \right) =\left[ \frac{%
\cos \left( \Delta \Theta \right) }{\sqrt{1-\sin \left( \Delta \Theta _{\max
}\right) }}\right] ,  \tag{2.17}
\end{equation}%
where the brackets indicate integer part. Also, one can put 
\begin{equation}
\Delta \Theta _{\max }\simeq \frac{C}{A}+O\left[ \left( \frac{C}{A}\right)
^{3}\right] .  \tag{2.18}
\end{equation}%
Thus, one can use a \textit{linear} approximation for $\Delta \Theta _{\max
} $ suitable for chaotic motions arising away from the limiting case of
tangency between the stable and unstable manifolds $\left( C/A\ll 1\right) $%
. It is worth mentioning that the last inequality is usually associated with
the observation of steady chaos (strange chaotic attractor).

\subsection{Enhancement of chaos}

It has been mentioned above that the mechanism for suppressing homoclinic
(and heteroclinic) chaos is the frustration of a homoclinic/heteroclinic
bifurcation, which prevents the appearance of horseshoes in the dynamics.
Chac\'{o}n (2001\textit{a}) showed that the enhancement of the initial chaos
is achieved by moving the system from the homoclinic tangency condition 
\textit{even more} than in the initial situation with no second periodic
excitation. Thus, enhancement of chaos can mean increasing the duration of a
chaotic transient, passing from transient to steady chaos, or increasing the
maximal Lyapunov exponent. Consider again that the family of systems modeled
by (1.1) satisfies the requirements of the Melnikov method. Similarly to the
preceding discussion of the suppression of chaos, one can assume any
particular form of the Melnikov function (2.4) to discuss the enhancement of
chaos. Therefore, consider, for instance, the following nonlinearly damped,
biharmonically driven, two-well Duffing oscillator:%
\begin{equation}
\overset{..}{x}-x+\beta x^{3}=-\delta \overset{.}{x}\left\vert \overset{.}{x}%
\right\vert ^{n-1}+F\cos \left( \omega t\right) -\eta \beta x^{3}\cos \left(
\Omega t+\Theta \right) ,  \tag{2.19}
\end{equation}%
where $\eta ,\Omega ,$ and $\Theta $ are the normalized amplitude factor,
frequency, and initial phase, respectively, of the chaos-controlling
parametric excitation $\left( 0<\eta \ll 1\right) $, and $\beta ,\delta
,n,F, $ and $\omega $ are the normalized parameters of the potential
coefficient, damping coefficient, damping exponent, chaos-inducing
amplitude, and chaos-inducing frequency, respectively $\left( 0<\delta ,F\ll
1,\beta >0,n=1,2,...\right) $. The application of the Melnikov method to
(2.19) yields the Melnikov function%
\begin{equation}
M^{\pm }(t_{0})=D\pm A\sin \left( \omega t_{0}\right) -C\sin \left( \Omega
t+\Theta \right) ,  \tag{2.20}
\end{equation}%
with%
\begin{eqnarray}
D &\equiv &-\delta \left( \frac{2}{\beta }\right) ^{\left( n+1\right)
/2}B\left( \frac{n+2}{2},\frac{n+1}{2}\right) ,  \notag \\
A &\equiv &\left( \frac{2}{\beta }\right) ^{1/2}\pi \omega F\func{sech}%
\left( \frac{\pi \omega }{2}\right) ,  \notag \\
C &\equiv &\frac{\pi \eta }{6\beta }\left( \Omega ^{4}+4\Omega ^{2}\right) 
\func{csch}\left( \frac{\pi \Omega }{2}\right) ,  \TCItag{2.21}
\end{eqnarray}%
where the positive (negative) sign refers to the right (left) homoclinic
orbit of the underlying integrable two-well Duffing oscillator ($\delta
=F=\eta =0$), and where $B\left( m,n\right) $ is the Euler beta function. It
has been demonstrated (Chac\'{o}n 2001\textit{a}) that, in general, a second
harmonic excitation can reliably play an enhancer or inhibitor role \textit{%
solely} from adjusting its initial phase. The Melnikov function $%
M^{+}(t_{0}) $ will be used here to illustrate the approach to the
enhancement of chaos. Indeed, consider that, in the absence of any second
parametric excitation $\left( C=0\right) $, the associated Melnikov function 
$M_{0}^{+}(t_{0})=-\left\vert D\right\vert +A\sin \left( \omega t_{0}\right) 
$ changes sign at some $t_{0}$, i.e., $\left\vert D\right\vert \leqslant A$.
If one now lets the second excitation act on the system such that $%
C\leqslant A-\left\vert D\right\vert $, this relationship represents a
sufficient condition for $M^{+}(t_{0})$ to change sign at some $t_{0}$.
Thus, a necessary condition for $M^{+}(t_{0})$ to always have the same sign $%
\left( M^{+}(t_{0})<0\right) $ is $C>A-\left\vert D\right\vert \equiv
C_{\min }$. It was above mentioned (Chac\'{o}n 1999) that a sufficient
condition for $C>C_{\min }$ to also be a sufficient condition for inhibiting
chaos is $\Omega =p\omega $ (subharmonic resonance condition), $C\leqslant
C_{\max }\equiv A/p^{2}$, $p$ an integer, and that $M_{0}^{+}(t_{0})$ and $%
-C_{\min ,\max }\sin \left( \Omega t_{0}+\Theta \right) $ are \textit{in
opposition}. This condition yields the optimal suppressory values $\Theta
_{opt}^{\sup }\equiv \Theta _{opt}$. It was demonstrated (Chac\'{o}n 2001%
\textit{a}) that imposing $M_{0}^{+}(t_{0})$ to be \textit{in phase} with $%
-C_{\min ,\max }\sin \left( \Omega t_{0}+\Theta \right) $ is a \textit{%
sufficient} condition for $M^{+}(t_{0})$ change sign at some $t_{0}$. This
condition provides the optimal enhancer values of the initial phase, $\Theta
_{opt}^{enh}$, in the sense that $M^{+}(t_{0})$ presents its highest maximum
at $\Theta _{opt}^{enh}$, i.e., one obtains the maximum gap from the
homoclinic tangency condition. Now, the following remarks are in order.

First, for a given homoclinic orbit forming (part of) a separatrix, one has
in general (i.e., for any Melnikov function (2.4)) that 
\begin{equation}
\left\vert \Theta _{opt}^{\sup }-\Theta _{opt}^{enh}\right\vert =\pi , 
\tag{2.22}
\end{equation}%
for each resonance order.

Second, for $C=C_{\min }$ there always exists a \textit{maximum-range}
interval%
\begin{equation}
\left[ \Theta _{opt}^{enh}-\Delta \Theta _{\max }^{enh}\left( C=C_{\min
}\right) ,\Theta _{opt}^{enh}+\Delta \Theta _{\max }^{enh}\left( C=C_{\min
}\right) \right]   \tag{2.23}
\end{equation}%
of permitted initial phases for enhancement of chaos in the sense that, for
values of $\Theta $ belonging to that interval, the maxima of $M^{+}(t_{0})$
are higher than those of $M_{0}^{+}(t_{0})$. Similarly, for $C=C_{\max }$
there always exists a \textit{different }maximum-range interval 
\begin{equation}
\left[ \Theta _{opt}^{enh}-\Delta \Theta _{\max }^{enh}\left( C=C_{\max
}\right) ,\Theta _{opt}^{enh}+\Delta \Theta _{\max }^{enh}\left( C=C_{\max
}\right) \right]   \tag{2.24}
\end{equation}%
of allowed initial phases for enhancement of chaos, and also%
\begin{equation}
\Delta \Theta _{\max }^{enh}\left( C=C_{\max }\right) \geqslant \Delta
\Theta _{\max }^{enh}\left( C=C_{\min }\right) ,  \tag{2.25}
\end{equation}%
which is a consequence of the dissipation. It must be emphasized that the
definition of $\Theta _{opt}^{enh}$ is general; i.e., it refers to any
resonance and any Melnikov function (2.4).

Third, for general separatrices, i.e., those formed by several homoclinic
and (or) heteroclinic loops, the above scenario of control of chaos holds
for \textit{each} \ homoclinic (heteroclinic) orbit. However, it is common
to find that the different homoclinic (heteroclinic) orbits of a given
separatrix yield \textit{distinct} $\Theta _{opt}^{enh}$ values. This is a
consequence of the survival of the symmetries existing in the absence of the
second excitation. Thus, the actual scenario is usually more complicated.
For instance, let $\Theta _{opt,r}^{\sup }$, $\Theta _{opt,l}^{\sup }$ be
the optimal values associated with the right and left homoclinic orbits,
respectively, of a typical separatrix with a \textquotedblleft
figure-of-eight\textquotedblright\ loop, as in the two-well Duffing
oscillator (2.19). One then obtains that the best chance for enhancing chaos
should now be at $\Theta _{opt}^{enh}\sim \left( \Theta
_{opt,r}^{enh}-\Theta _{opt,l}^{enh}\right) /2\,\func{mod}\left( 2\pi
\right) .$ See Chac\'{o}n (2001\textit{a}) for more details.

\subsection{Further developments}

The case of subharmonic resonance between the chaos-inducing and
chaos-controlling frequencies has been briefly discussed above. However, a
number of theoretical (Salerno 1991; Salerno \& Samuelsen 1994), numerical
(Braiman \& Goldhirsch 1991), and experimental (Uchida \textit{et al.} 1998)
studies show that chaos can be reliably controlled by other non-subharmonic
resonances. The work of Chac\'{o}n (2001\textit{b}) presents a
Melnikov-method-based approach concerning reduction of homoclinic and
heteroclinic instabilities for the family of systems (1.1) where the
harmonic excitations verify an ultrasubharmonic resonance condition: $\Omega
/\omega =p/q$, $q>1\left( p\neq q\right) $, $p,q$ positive integers and $%
\Omega \left( \omega \right) $ the chaos-suppressing (inducing) frequency.
Such general results can be used to approach the case of \textit{%
incommensurate} chaos-suppressing excitations by means of a series of ever
better rational approximations, which are the successive convergents of the
infinite continued fraction associated with the irrational ratio $\Omega
/\omega $. This procedure has been much employed in characterizing strange
non-chaotic attractors in quasiperiodically forced systems as well as in
studying phase-locking phenomena in both Hamiltonian and dissipative
systems. To illustrate the method one intentionally chooses the golden
section $\Omega /\omega =\Phi \equiv \left( \sqrt{5}-1\right) /2$, since it
is the irrational number which is the worst approximated by rational numbers
(in the sense of the size of the denominator). As is well-known, the golden
section can be approximated by the sequence of rational numbers $\left(
\Omega /\omega \right) _{i}=F_{i-1}/F_{i}$ where $F_{i}=1,1,2,3,5,8,...$,
are the Fibonacci numbers such that $\lim_{i\rightarrow \infty }\left(
\Omega /\omega \right) _{i}=\Phi $. For each $\left( \Omega /\omega \right)
_{i}$ one replaces each quasiperiodically excited system 
\begin{equation}
\overset{..}{x}+\frac{dU(x)}{dx}=-d(x,\overset{.}{x})+p_{c}(x,\overset{.}{x}%
)har(\omega t)+p_{s}\left( x,\overset{.}{x}\right) har^{\prime }\left( \Phi
\omega t+\Psi _{h,h^{\prime }}\right)   \tag{2.26}
\end{equation}%
by the respective periodically excited system%
\begin{equation}
\overset{..}{x}+\frac{dU(x)}{dx}=-d(x,\overset{.}{x})+p_{c}(x,\overset{.}{x}%
)har(\omega t)+p_{s}\left( x,\overset{.}{x}\right) har^{\prime }\left( \frac{%
F_{i-1}}{F_{i}}\omega t+\Psi _{h,h^{\prime }}\right)   \tag{2.27}
\end{equation}%
giving a sequence of periodically excited systems whose associated
frequencies satisfy an ultrasubharmonic resonance condition. The work of Chac%
\'{o}n \& Mart\'{\i}nez (2002) applied this approach to the problem of the
reduction of chaotic escape from a potential well using the simple model%
\begin{equation}
\overset{..}{x}=x-\beta x^{2}-\delta \overset{.}{x}+\gamma \sin \left(
\omega t\right) -\beta \eta x^{2}\sin \left( \Omega t+\Theta \right) , 
\tag{2.28}
\end{equation}%
where $\beta \eta x^{2}\sin \left( \Omega t+\Theta \right) $ is the
escape-suppressing excitation. They found that, for irrational
escape-suppressing frequencies, the effective escape-reducing initial phases
are found to lie close to the \textit{accumulation} points of the set of
suitable initial phases that are associated with the complete series of
convergents up to the convergent giving the chosen rational approximation.

A Melnikov-method-based approach (Chac\'{o}n 2002) was presented concerning
the \textit{relative effectiveness }of harmonic excitations in suppressing
homoclinic (and heteroclinic) chaos of the family (1.1) for the main
resonance between the chaos-inducing and chaos-suppressing excitations. A
criterion based on the area in the suppressory amplitude/initial phase
parameter plane, where suppression of homoclinic chaos is guaranteed, was
deduced and shown to be useful in choosing the most suitable of the possible
chaos-suppressing excitations. Additionally, the choice of the most suitable
chaos-suppressing excitation was shown to exhibit \textit{sensitivity }to
the particular initial chaotic state.

The work of Chac\'{o}n \textit{et al.} (2003) presents general findings
concerning control of chaos for the family 
\begin{equation}
\overset{..}{x}+\frac{dU(x)}{dx}=-d\left( x,\overset{.}{x}\right)
+\sum_{i=1}^{N}h_{ch,i}\left( x,\overset{.}{x}\right)
F_{ch,i}(t)+\sum_{j=1}^{M}h_{co,j}\left( x,\overset{.}{x}\right) F_{co,j}(t),
\tag{2.29}
\end{equation}%
where $U(x)$ is a general potential, $-d\left( x,\overset{.}{x}\right) $
represents a generic dissipative force, $\sum_{i=1}^{N}h_{ch,i}\left( x,%
\overset{.}{x}\right) F_{ch,i}(t)$ is a general multiple chaos-inducing
excitation, and $\sum_{j=1}^{M}h_{co,j}\left( x,\overset{.}{x}\right)
F_{co,j}(t)$ is an as yet undetermined suitable multiple chaos-controlling
excitation, with $F_{ch,i}(t),\,F_{co,j}(t)$ being harmonic functions of
common frequency $\omega $ and initial phases $0$ ($i=1,...,N$), $\varphi
_{j}$ ($j=1,...,M$). The effectiveness of this approach in suppressing
spatio-temporal chaos of chains of identical chaotic coupled oscillators was
demonstrated through the example of coupled Duffing oscillators, where
coherent oscillations were achieved under \textit{localized} control.

The work of Chac\'{o}n \textit{et al. }(2002) studied the robustness of the
suppression of bidirectional chaotic escape of a harmonically driven
oscillator from a quartic potential well by the application of weak
parametric excitations. It was numerically shown that Melnikov-method-based
theoretical predictions also work for harmonic escape-inducing excitations
in the presence of external noise, and for chaotic-escape-inducing
excitations having a sharp Fourier component with a sufficiently high power.

The method proposed in the work of Lenci \& Rega (2003) consists of choosing
the shape of external and/or parametric periodic excitations, which permits
one avoid, in an optimal manner, a homoclinic bifurcation. They numerically
investigated the effectiveness of the control method with respect to the
basin erosion and escape phenomena of a perturbed Helmholtz oscillator.

\section{Some applications}

\subsection{Taming chaotic escape from a potential well}

The work of Chac\'{o}n \textit{et al.} (1996, 1997, 2001) and Balibrea 
\textit{et al.} (1998) applies the above Melnikov-method-based approach to
the problem of chaotic escape from a potential well. This is a general and
ubiquitous phenomenon in science. Indeed, one finds it in very distinct
contexts: the capsizing of a boat subjected to trains of regular waves
(Thompson 1989), the stochastic escape of a trapped ion induced by a
resonant laser field (Chac\'{o}n \& Cirac, 1995), and the escape of stars
from a stellar system (Contopoulos \textit{et al}. 1993) are some important
examples. Remarkably, such complex escape phenomena can often be well
described by a low-dimensional system of differential equations. The case
considered by Chac\'{o}n and coworkers is that where escape is induced by an
external periodic excitation added to the model system, so that, before
escape, chaotic transients of unpredictable duration (due to the fractal
character of the basin boundary) are usually observed for orbits starting
from chaotic generic phase space regions (such as those surrounding
separatrices), in both dissipative and Hamiltonian systems. In particular,
Chac\'{o}n \textit{et al.} (1996) studied the simplest model for a universal
chaotic escape situation:%
\begin{equation}
\overset{..}{x}-x+\beta x^{2}=-\delta \overset{.}{x}+\gamma \sin \left(
\omega t\right) +\binom{-\beta \eta x^{2}\sin \left( \Omega t+\Theta \right) 
}{\beta \eta x\sin \left( \Omega t+\Theta \right) },  \tag{3.1}
\end{equation}%
where $\beta \eta x^{2}\sin \left( \Omega t+\Theta \right) $ and $\beta \eta
x\sin \left( \Omega t+\Theta \right) $ are the (independently considered)
escape-suppressing parametric excitations. It was demonstrated that the
parametric excitation of the linear (quadratic) term suppress chaotic escape
more efficiently than that of the quadratic (linear) term for small (large)
driving periods of the primary chaos-inducing excitation. Chac\'{o}n \textit{%
et al}. (1997) studied the inhibition of chaotic escape of a driven
oscillator from the cubic potential well that typically models a metastable
system close to a fold:%
\begin{equation}
\overset{..}{x}+x-\beta x^{2}=-\binom{\delta _{1}\overset{.}{x}}{\left(
\delta _{2}x^{2}+\delta _{3}x^{4}\right) \overset{.}{x}}+\gamma \cos \left(
\omega t\right) -\eta x\cos \left( \Omega t+\Theta \right) ,  \tag{3.2}
\end{equation}
where $\delta _{1}\overset{.}{x}$ and $\left( \delta _{2}x^{2}+\delta
_{3}x^{4}\right) \overset{.}{x}$ are the (independently considered) linear
and nonlinear damping terms, respectively. They demonstrated that the
effectiveness of a parametric excitation at suppressing chaotic escape from
such a cubic potential well diminishes as the system approaches a period-1 
\textit{parametric resonance}, and that, for linear damping, the parametric
excitation inhibits chaotic escape more efficiently than for nonlinear
damping. The role of a nonlinear damping term, proportional to the \textit{n}%
th power of the velocity, on the escape-inhibition scenario is considered in
the work of Chac\'{o}n \textit{et al.} (2001):%
\begin{equation}
\overset{..}{x}+x-x^{2}=-\delta \overset{.}{x}\left\vert \overset{.}{x}%
\right\vert ^{n-1}+\gamma \cos \left( \omega t\right) +\eta x^{2}\cos \left(
\Omega t+\Theta \right) ,  \tag{3.3}
\end{equation}%
where $\eta x^{2}\cos \left( \Omega t+\Theta \right) $ is the
escape-suppressing parametric excitation. In this case, the effectiveness of
the parametric excitation of the quadratic potential well at inhibiting
chaotic escape diminishes as the system approaches either a period-1 or a
period-2 parametric resonance. Also, the effectiveness of the parametric
excitation in the presence of the nonlinear dissipative force is less than
for a linear dissipative force.

\subsection{Taming chaotic solitons in Frenkel-Kontorova chains}

Control of chaos in spatially extended systems is one of the most important
and challenging problems in the field of nonlinear dynamics. Instances of
possible applications include the stabilization of superconducting
Josephson-junction arrays (Barone \& Paterno 1982), periodic patterns in
optical turbulence, and semiconductor laser arrays (Sch\"{o}ll 2001), to
cite just a few. Mart\'{\i}nez \& Chac\'{o}n (2004) presented a
Melnikov-method-based general theoretical approach to control chaotic 
\textit{solitons} in damped, noisy and driven Frenkel-Kontorova chains.
Specifically, they studied the model%
\begin{eqnarray}
\overset{..}{x}_{j}+\frac{K}{2\pi }\sin \left( 2\pi x_{j}\right)
&=&x_{j+1}-2x_{j}+x_{j-1}-\alpha \overset{.}{x}_{j}+F\cos \left( \omega
t\right)  \notag \\
&&+\beta F\cos \left( \Omega t+\varphi \right) +\xi \left( t\right) , 
\TCItag{3.4}
\end{eqnarray}%
where $\beta F\cos \left( \Omega t+\varphi \right) $ is the
chaos-suppressing excitation, and $\xi \left( t\right) $ is a bounded noise
term. They obtained an effective equation of motion governing the dynamics
of the soliton center of mass for which they deduced Melnikov-method-based
predictions concerning the regions in the control parameter space where
homoclinic bifurcations are frustrated. Numerical simulations indicated that
such theoretical predictions can be reliably applied to the original
Frenkel-Kontorova chains, even for the case of \textit{localized}
application of the soliton-taming excitations. It is worth mentioning that
the same effectiveness of such a localized control in suppressing
spatio-temporal chaos of chains of identical chaotic coupled oscillators was
demonstrated through the example of coupled Duffing oscillators (Chac\'{o}n 
\textit{et al.} 2003).

\subsection{Taming chaotic charged particles in the field of an
electrostatic wave packet}

The interaction of charged particles with an electrostatic wave packet is a
basic and challenging problem appearing in diverse fundamental fields such
as astrophysics, plasma physics, and condensed matter physics. While the
Hamiltonian approach to this problem is suitable in many physical contexts,
the consideration of dissipative forces seems appropriate in diverse
phenomena such as the stochastic heating in the dynamics of charged
particles interacting with plasma oscillations. In any case, stochastic
(chaotic) dynamics already appears (can appear) when the wave packet solely
consists of two electrostatic plane waves. Such a non-regular behavior of
the charged particles may yield undesirable effects on a number of
technological devices such as the destruction of magnetic surfaces in
tokamaks. Thus, apart from its general intrinsic interest, the problem of
regularization of the dissipative dynamics of charged particles in an
electrostatic wave packet by a small-amplitude uncorrelated wave (which is
added to the initial wave packet) is especially relevant in plasma physics.
Chac\'{o}n (2004) considered the simplest model equation to examine this
problem:%
\begin{eqnarray}
\overset{..}{x}+\delta \overset{.}{x} &=&-\frac{e}{m}\left[ E_{0}\sin \left(
k_{0}x-\omega _{0}t\right) +E_{c}\sin \left( k_{c}x-\omega _{c}t\right) %
\right]  \notag \\
&&-\frac{e}{m}E_{s}\sin \left( k_{s}x-\omega _{s}t\right) ,  \TCItag{3.5}
\end{eqnarray}%
where $E_{c}\sin \left( k_{c}x-\omega _{c}t\right) $ and $E_{s}\sin \left(
k_{s}x-\omega _{s}t\right) $ are the chaos-inducing and chaos-suppressing
waves, respectively. In a reference frame moving along the main wave $%
E_{0}\sin \left( k_{0}x-\omega _{0}t\right) ,$ (3.5) transforms into a
perturbed pendulum equation which is capable of being studied by means of
Melnikov%
%TCIMACRO{\U{b4}}%
%BeginExpansion
\'{}%
%EndExpansion
s method. Two suppressory mechanisms were identified: One mechanism requires
chaos-inducing and chaos-suppressing waves to have both commensurate
wavelengths and commensurate relative phase velocities, while the other
allows chaos to be tamed when these quantities are incommensurate.

\section{Conclusions and open problems}

The present review summarizes some of the main results and applications of a
preliminary theoretical approach to control chaos in dissipative,
non-autonomous dynamical systems, capable of being studied by Melnikov%
%TCIMACRO{\U{b4}}%
%BeginExpansion
\'{}%
%EndExpansion
s method, by means of periodic excitations. Diverse extensions and
applications of the current theory remain to be developed. Among them:

(i) To obtain the boundaries of the regularization regions in the control
parameter space for the case of a general resonance (not just the main)
between the involved excitations.

(ii) To extend the theoretical approach to (some family of) multidimensional
systems capable of being studied by (some generalized version of) Melnikov%
%TCIMACRO{\U{b4}}%
%BeginExpansion
\'{}%
%EndExpansion
s method.

(iii) To develop a multiharmonic control theory beyond the main resonance
case.

(iv) To extend the theoretical approach for the case of periodic excitations
to the case of random excitations.

(v) To obtain analytical approximations of the regularized responses for the
deterministic case of a general resonance between the chaos-inducing and
chaos-suppressing excitations.

(vi) To extend the current theory described for harmonic excitations to the
case of general periodic excitations (both chaos-inducing and
chaos-controlling). In particular, the \textit{waveform effect} should be
taken into account in the control scenario.

(vii) To extend the current theory to the case where the chaos-controlling
excitation is a parametric excitation of the amplitude of the chaos-inducing
excitation, as well as to the case where it is a parametric excitation of
the frequency of the chaos-inducing excitation.

(viii) To apply the current theory to ratchet systems to improve the
directed energy transport.

(ix) To apply the current theory to control chaotic population oscillations
between two coupled Bose-Einstein condensates with time-dependent asymmetric
potential and damping.

\begin{acknowledgement}
The author acknowledges financial support from Spanish MCyT and European
Regional Development Fund (FEDER) program through BFM2002-00010 project.
\end{acknowledgement}

\end{document}